\documentclass[conference]{IEEEtran}
\IEEEoverridecommandlockouts
\usepackage{cite}
\usepackage{times,amsmath,amssymb,psfrag,stfloats}
\usepackage{tabularx}
\usepackage{graphicx}
\usepackage{epsfig}
\usepackage{epstopdf}
\graphicspath{{./pics/}} \DeclareGraphicsExtensions{.pdf,.jpg,.png,.eps}
\usepackage{color}
\usepackage{algorithmic}
\usepackage{algorithm}
\usepackage{amsmath}
\usepackage{enumitem}
\usepackage{multirow}
\usepackage{flushend}
\usepackage{fancyhdr}
\usepackage[bookmarks=false]{hyperref}

\fancypagestyle{firstpage}{%
 \fancyhf{}%
}

\makeatletter

\newcommand{\newlineauthors}{%
\end{@IEEEauthorhalign}\hfill\mbox{}\par
\mbox{}\begin{@IEEEauthorhalign}}  

\renewcommand{\baselinestretch}{0.97}

\begin{document}

\title{Exploration of Multi-Element Collaborative Research and Application for Modern Power System Based on Generative Large Models}

\author{\IEEEauthorblockN{Lu Cheng\IEEEauthorrefmark{1}\IEEEauthorrefmark{2},
Qixiu Zhang\IEEEauthorrefmark{3}, Beibei Xu\IEEEauthorrefmark{4}, Zhiwei Huang\IEEEauthorrefmark{5}, Cirun Zhang\IEEEauthorrefmark{4}, Yanan Lyu\IEEEauthorrefmark{8}, and Fan Zhang\IEEEauthorrefmark{6}}
\IEEEauthorblockA{\IEEEauthorrefmark{2}Bay Area Founders Club, INC., US \& Georgia Institute of Technology, US}
\IEEEauthorblockA{\IEEEauthorrefmark{3}RIS Department, Roche.Inc, US}
\IEEEauthorblockA{\IEEEauthorrefmark{4}Pratt School of Engineering, Duke University, US}
\IEEEauthorblockA{\IEEEauthorrefmark{5}Yellow River Conservancy Technical Institute, Henan, China \& Tianjin University, China}
\IEEEauthorblockA{\IEEEauthorrefmark{8}Monetization Department, Meta, US}
\IEEEauthorblockA{\IEEEauthorrefmark{6}Transmission Ops \& Planning, Electric Power Research Institute, US}
\IEEEauthorrefmark{1}{Corresponding Author: lcheng311@gatech.edu \& joyce@bayareafoundersclub.com}}

%\author{\IEEEauthorblockN{Yanan Lyu}
%\IEEEauthorblockA{\textit{Meta, US}\\ 
%Meta, US}
%\and
%\IEEEauthorblockN{Qixiu Zhang*}
%\IEEEauthorblockA{\textit{RIS Department, Roche.Inc, US}\\
%zhangqiuxiumum@gmail.com}
%*Corresponding author
%\and
%\IEEEauthorblockN{Cirun Zhang}
%\IEEEauthorblockA{\textit{Pratt School of Engineering,}\\ 
%Duke University, US}
%\newlineauthors
%\IEEEauthorblockN{Zhiwei Huang}
%\IEEEauthorblockA{\textit{School of Mechanical Engineering,}\\ 
%Tianjin University, China}
%\and
%\IEEEauthorblockN{Fan Zhang}
%\IEEEauthorblockA{\textit{Transmission Ops \& Planning,}\\ 
%Electric Power Research Institute, US}
%\and
%\IEEEauthorblockN{Beibei Xu}
%\IEEEauthorblockA{\textit{Pratt School of Engineering,}\\ 
%Duke University, US}
%}

\maketitle
\thispagestyle{firstpage}

\begin{abstract}
The transition to intelligent, low-carbon power systems necessitates advanced optimization strategies for managing renewable energy integration, energy storage, and carbon emissions. Generative Large Models (GLMs) provide a data-driven approach to enhancing forecasting, scheduling, and market operations by processing multi-source data and capturing complex system dynamics. This paper explores the role of GLMs in optimizing load-side management, energy storage utilization, and electricity carbon, with a focus on Smart Wide-area Hybrid Energy Systems with Storage and Carbon (SGLSC). By leveraging spatiotemporal modeling and reinforcement learning, GLMs enable dynamic energy scheduling, improve grid stability, enhance carbon trading strategies, and strengthen resilience against extreme weather events. The proposed framework highlights the transformative potential of GLMs in achieving efficient, adaptive, and low-carbon power system operations.
\end{abstract}


\section{Introduction} 
As the global energy landscape transitions toward sustainability and carbon neutrality, the modernization of power systems has become a critical priority \cite{10252793}. A modern power system is characterized by the dominance of renewable energy sources, the coordinated interaction among generation, grid, load, and storage, and the deep integration of electricity and carbon management. The objective is to establish a secure, stable, intelligent, and low-carbon power infrastructure. However, the inherent intermittency and uncertainty of renewables, increasing operational complexity, and the challenges of multi-agent optimization render traditional dispatch and control methods insufficient \cite{10838247,9299496,8999526}.

Generative large models (GLMs), leveraging advanced deep learning techniques, offer a transformative approach to power system intelligence \cite{9726814}. These models excel in data-driven learning, high-dimensional pattern generation, and complex system optimization. Their core technological components include multi-modal data fusion, large-scale time-series forecasting, intelligent scheduling, reinforcement learning, and causal inference. By integrating these capabilities, GLMs facilitate multi-factor coordinated optimization across generation, grid operation, demand response, energy storage, and carbon emission management, enhancing the efficiency and adaptability of power systems \cite{9721662}.

This paper explores the application of GLMs in modern power systems, focusing on their role in renewable energy forecasting, generation-side optimization, intelligent dispatch and stability control, demand response, energy storage regulation, and carbon emission forecasting. Additionally, it examines their potential for multi-factor optimization and discusses the challenges and future directions for their practical implementation. This study aims to provide theoretical insights and technical guidance for the development of intelligent, resilient, and sustainable power systems.

\section{Research and Application of GLMs on Generation}
The generation side plays a critical role in ensuring power system stability and efficiency. With renewable sources such as wind and photovoltaic (PV) power becoming dominant, their variability and uncertainty—driven by meteorological conditions—pose significant operational challenges. Addressing these challenges requires enhanced forecasting accuracy and optimized dispatch strategies. GLMs leverage large-scale data patterns and integrate heterogeneous information, such as meteorological data, grid operation records, and historical generation data, to enhance renewable energy forecasting capabilities and optimize dispatch decisions.

\subsection{Application in Renewable Energy Forecasting}
The traditional forecasting of renewable energy sources $P_t$ at time $t$ depends on multiple meteorological and system variables as (\ref{eq1}) \cite{10166280}
\begin{equation} 
P_t = f(W_t, S_t, T_t,H_t)+\varepsilon_t
\label{eq1} 
\end{equation}
\noindent where $Wt$ represents wind speed and direction, $S_t$ denotes solar radiation intensity, $T_t$ is the ambient temperature, $H_t$ includes historical power generation data, $\varepsilon_t$ accounts for prediction errors. Traditional forecasting models struggle with capturing complex dependencies in $f(.)$, particularly under extreme conditions. GLMs enhance forecasting accuracy by integrating multi-modal data sources and employing deep learning architectures. ransformer-based generative models \cite{9462413,9555586,9869310}, as (\ref{eq2}),  for example, predict future generation by learning temporal dependencies:
 \begin{equation} 
 \hat{P}_{t+1;t+n} = GLM(P_{t-m,t},X_{t-m,t})
 \label{eq2} 
\end{equation}
 where $X_{t-m,t}$ represents exogenous inputs (e.g., weather conditions), and $\hat{P}_{t+1;t+n}$ is the predicted generation for the future time window.

\subsection{Generative Large Models for Power Generation Optimization and Dispatch}
Traditional power dispatch methods, reliant on physical models and optimization algorithms, face computational challenges and struggle with real-time uncertainty in renewable energy generation. GLMs enhance dispatch efficiency by leveraging data-driven learning, reinforcement learning, and intelligent optimization \cite{10512990}. They dynamically adjust power dispatch strategies while considering load demand, energy storage, and grid constraints. By predicting renewable output and integrating regulatory resources such as thermal and hydropower, GLMs optimize dispatch, reducing wind and solar curtailment. Additionally, through cloud-edge-terminal collaboration, they enable decentralized, real-time dispatch for distributed energy sources, improving flexibility and responsiveness \cite{9583953}.

\section{Research and Application of GLMs on Grid}

\begin{figure}[h!bt]
\centerline{\includegraphics[scale = 0.40] {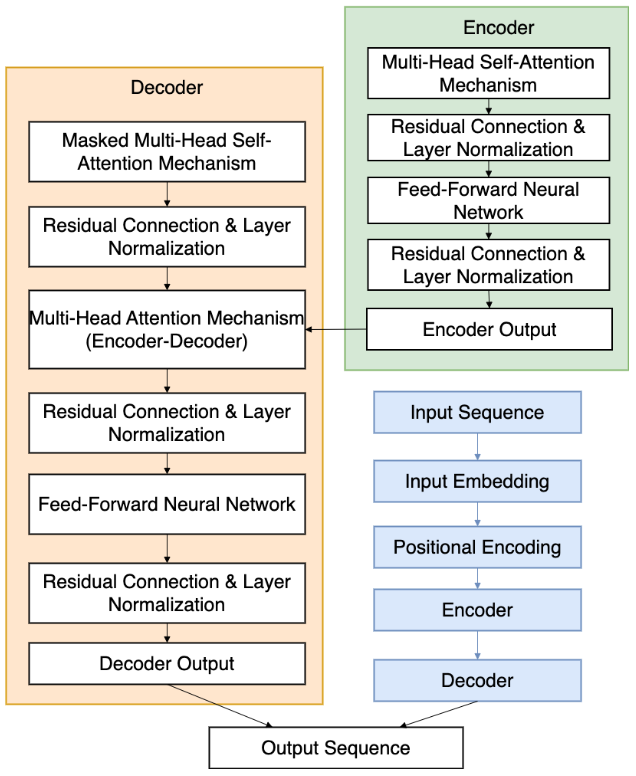}}
\caption{Structure of the Transformer model.}\label{fig1}
\end{figure}

GLMs can help meet the needs for real-time operation, flexibility, and intelligence in modern power grid by integrating various data, including grid operations, renewable energy forecasts, load data, and external environmental factors. By leveraging advanced techniques like time-series modeling, causal inference, and reinforcement learning, GLMs enhance grid situational awareness, optimize grid dispatch, and improve risk forecasting, contributing to better security, stability, and efficiency in grid operations.

In terms of situational awareness, traditional grid operation methods are limited by their reliance on physical models and expert experience, making it difficult to accurately predict variations in grid states \cite{10632043}. GLMs can overcome this limitation by integrating diverse data sources, such as SCADA, PMU, and meteorological data, to build comprehensive models that improve real-time awareness and enable the detection of abnormal events \cite{9337223}.

For grid-wide operational state prediction, GLMs using advanced deep learning architectures, such as the transformer models \cite{9321176} showns as Fig. \ref{fig1}, can accurately forecast grid states, including fluctuations in voltage, current, and power. This predictive capability helps to ensure grid stability by anticipating future conditions and enabling proactive responses.

When it comes to intelligent anomaly detection and fault early warning, GLMs integrate techniques like Generative Adversarial Networks (GANs) \cite{8667290} or Variational Autoencoders (VAEs) \cite{9119397} to learn from past data and detect abnormal patterns, such as voltage instability or frequency anomalies, at an early stage. This early detection can significantly enhance grid security by allowing for quicker interventions before failures escalate.

In the domain of grid security and risk forecasting, GLMs offer valuable advancements in fault diagnosis and grid restoration. Traditional methods of fault diagnosis depend on predefined rules, which may not be adaptable enough to handle the evolving scenarios presented by high renewable energy shares. Combing with graph neural networks (GNNs) \cite{10176302} and deep generative models \cite{9555209}, GLMs can automatically learn grid topologies, enabling more adaptive and accurate fault detection. By generating synthetic grid data under normal conditions and comparing it to real-time data, GLMs can detect anomalies and quickly identify faults \cite{8881675}. Additionally, GLMs can develop adaptive fault restoration strategies using reinforcement learning, which helps speed up recovery after faults and improves grid resilience.

Furthermore, GLMs play a crucial role in grid operation risk assessment. By integrating weather forecasts, historical grid data, and meteorological disaster information, GLMs can predict the impact of extreme weather events on grid operations and provide preventive measures. They can also build causal networks to analyze the propagation of risk in different operational strategies, which improves the ability of the network to manage complex failures and improve the resilience of the system \cite{9729853}.

\section{Research and Application of GLMs in Load}
The intelligent transformation of power systems is increasingly reliant on active regulation and optimization at the load side. With the growing integration of distributed energy resources, energy storage systems, demand response mechanisms, and electric vehicles (EVs), traditional forecasting and control approaches often struggle to meet the evolving needs for accurate scheduling, dynamic interaction, and improved energy efficiency \cite{10379576}. GLMs provide a promising solution by synthesizing multi-source data, including electricity consumption patterns, environmental variables, market signals, and renewable energy outputs. This enables advanced applications in load forecasting, demand response, intelligent energy management, and anomaly detection, contributing to a more adaptive, low-carbon, and resilient approach to load management.

\subsection{Generative Large Models for Load Forecasting and Demand Response}
Accurate load forecasting is essential for effective power system scheduling, directly influencing grid stability and operational efficiency \cite{9319813}. Conventional methods often face difficulties in capturing the complex, non-linear variations in load patterns, especially under extreme weather or fluctuating market conditions. GLMs using models such as transformers, diffusion models \cite{9698190}, and VAEs, can capture intricate temporal dependencies and simulate future load fluctuations with greater accuracy. In GLMs, a computational load forecasting function \cite{7741548} often involves neural networks, decision trees, or other machine learning models, which learn patterns in the input data to produce an accurate forecast.

For a more advanced formulation in the context of GLMs, this could also be represented as:
\begin{equation} 
L(t)= \sum_{i=1}^n\beta_iX_i(t)+\varepsilon
 \label{eq2} 
\end{equation}
\noindent where $X_i(t)$ represents various predictor variables, such as weather, historical load, or time-related features. $\beta_i$ are the coefficients determined by the GLM model during the training phase. $\varepsilon$ is the error term.

GLMs also offer a more dynamic approach to demand response (DR), a vital mechanism for balancing supply and demand while integrating renewable energy. Traditional DR strategies often use static models or rule-based heuristics, which not account for real-time variations in user behavior and grid conditions \cite{10368142}. GLMs, especially when combined with reinforcement learning, can dynamically model user participation and optimize response strategies that balance grid stability, economic incentives, and user comfort. This ability to generate adaptive, multi-objective DR solutions helps reduce peak loads, lower electricity costs, and improve renewable energy utilization. GLMs are particularly effective in managing EV charging patterns and optimizing vehicle-to-grid (V2G) interactions \cite{10166065}, enhancing overall load regulation efficiency.

\subsection{Intelligent Load Management and Anomaly Detection}
As distributed energy systems and storage technologies reshape consumption patterns, intelligent load management becomes crucial for improving energy efficiency \cite{10423192}. GLMs can learn historical consumption behaviors, forecast future demands, and optimize load distribution at both individual and regional levels. Applied across smart homes, industrial facilities, or city-wide infrastructures, GLMs support dynamic energy management strategies that improve self-consumption rates in photovoltaic systems and increase the financial returns of energy storage assets. By forecasting load fluctuations \cite{8978655} and electricity price trends \cite{10875413}, these models enable better scheduling of energy resources, ultimately reducing costs and boosting system resilience.

Another key area where GLMs prove valuable is in anomaly detection, particularly in identifying irregular consumption behaviors that may indicate electricity theft, equipment failures, or inefficiencies. Traditional detection systems, often rule-based, struggle with evolving consumption patterns, resulting in delayed or inaccurate diagnoses \cite{8683792}. GLMs can utilize GANs to establish baseline consumption profiles, that can more effectively detect deviations, allow for quicker identification of abnormal behaviors. In addition, VAEs and time series generation models help in early fault detection in electrical devices by analyzing patterns from load data. This supports predictive maintenance strategies, preventing costly equipment failures. GLMs also offer information on energy efficiency improvements by simulating the effects of various optimization measures, helping users adopt strategies to minimize waste and enhance energy performance \cite{6246270}.

\section{Application and Research of GLM in Energy Storage}

\begin{figure}
\centerline{\includegraphics[scale = 0.19] {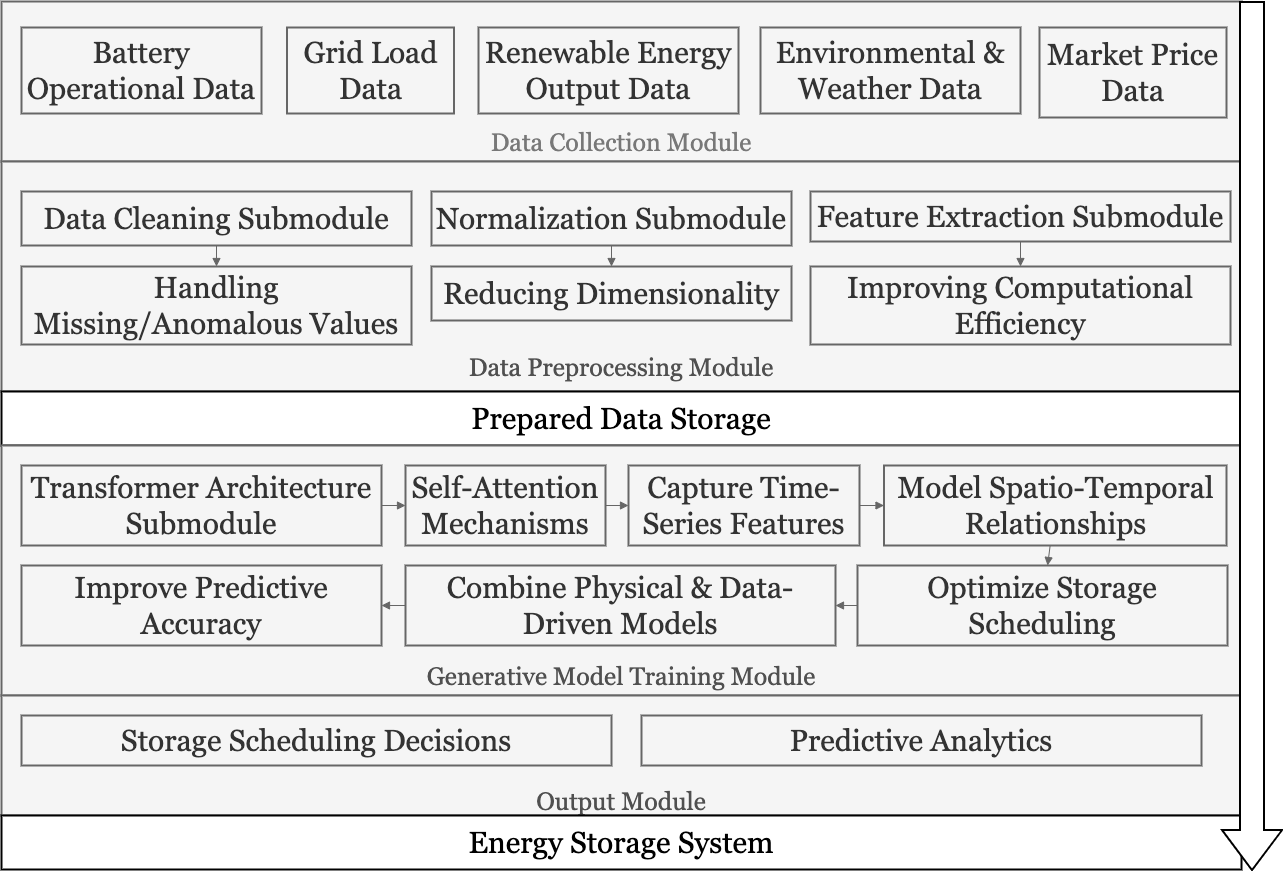}}
\caption{Framework for GLMs in Energy Storage System}\label{fig3}
\end{figure}

Energy storage plays a crucial role in mitigating the volatility of renewable energy generation and enhancing grid flexibility, making it an indispensable technology in modern power systems. However, issues such as optimization of storage scheduling, charging and discharge strategies, and evaluating health status remain challenging \cite{7491338}. GLMs offer a promising solution as Fig.\ref{fig3}. Due to their powerful data analysis capabilities, multimodal learning abilities, and adaptive optimization features, GLMs can provide new pathways for intelligent management and optimization of energy storage systems.

GLMs have demonstrated considerable potential in various aspects of energy storage management. One of the key applications is integrating multimodal data, such as battery operation data, environmental conditions, and load data, into a unified model \cite{10838247}. GLMs excel at extracting critical features from these diverse data sources and constructing efficient, data-driven models. This integration is particularly valuable for system states monitoring, where generative models can create highly accurate predictive models for battery health. By utilizing both historical and real-time operational data, these models enable the early detection of battery degradation and faults, facilitating timely maintenance strategies and prolonging the lifespan of storage systems.

In addition to health monitoring, GLMs can optimize charging and discharging strategies by combining real-time grid scheduling demands with renewable energy forecasts \cite{10398613}. This integration enhances energy utilization efficiency, ensuring that energy storage systems operate at their full potential. GLMs are also valuable in managing storage during extreme weather conditions, such as hurricanes or heavy rainfall, by providing precise response strategies that ensure grid stability and reliable power supply. Furthermore, these models can optimize discharge timings based on electricity price forecasts, helping maximize economic returns from energy storage systems.

Regarding storage scheduling optimization, GLMs are capable of supporting scheduling across multiple timeframes. Short-term scheduling (minute-level) optimizes real-time charging and discharging decisions, as well as battery health management \cite{8993691}. Mid-term scheduling (hourly) helps develop strategies that consider power market prices, while long-term scheduling (daily or weekly) focuses on predicting future load demands and optimizing storage capacity and expansion \cite{10379573}. These models also support grid dispatch, contributing to real-time grid operation optimization by integrating energy storage with load-side management. Additionally, they enhance demand-side response capabilities, increasing grid flexibility through intelligent control. In extreme weather scenarios, generative models assist in formulating early storage strategies to alleviate potential grid pressures.

\section{ Research and Application of GLMs in Electricity Carbon}
Managing electricity carbon is a critical component in developing a green, low-carbon power system, encompassing aspects such as carbon emission monitoring, carbon footprint tracking, and optimization of carbon market trading. However, the complexity of power system carbon emissions data—due to its diverse sources, long time spans, and intricate interdependencies—makes precise modeling and optimization challenging using traditional methods \cite{10252793,10512421}. GLMs with their advanced data processing and pattern recognition capabilities, can effectively integrate multi-source data, establishing an intelligent electricity carbon management system to support carbon emission forecasting, carbon quota optimization, and carbon trading decision-making.

GLMs provide significant value in managing carbon emissions by constructing high-precision prediction systems that integrate power generation and consumption data. These predictive models assist policymakers and businesses in formulating effective carbon management strategies. By leveraging multi-modal data, they track carbon footprints throughout the power supply chain  \cite{9721566}, enhancing transparency and facilitating green power certification.

In carbon markets, GLMs forecast carbon price trends and optimize trading strategies, ensuring efficient allocation of carbon assets. They also support the integration of renewable energy sources into the grid by dynamically adjusting carbon optimization models to minimize emissions. During extreme weather events, these models simulate variations in power-related carbon emissions under different meteorological conditions \cite{9056463}, providing valuable insights for scheduling and optimization.

GLMs enable predictions and optimization across multiple time horizons. In the short term (minutes to hours), they monitor real-time carbon emissions, supporting dynamic grid scheduling adjustments. Mid-term forecasts (days to weeks) predict carbon emission trends, guiding carbon trading strategies and policy decisions. Long-term predictions (months to years) assist in the planning of low-carbon power system development \cite{10512421}, ensuring alignment with regulatory and environmental goals. 

These models can also be applied in carbon market operations and grid dispatch. By optimizing carbon quota distribution based on generation plans, they facilitate greater adoption of clean energy while minimizing emissions. Intelligent scheduling strategies derived from GLMs further enhance carbon reduction efforts by balancing economic and environmental considerations in power system operations.

\section{Interaction Analysis and Application Scenarios of Source-Grid-Load-Storage-Carbon(SGLSC)}

\begin{figure*}[h!bt]
\centerline{\includegraphics[scale = 0.165] {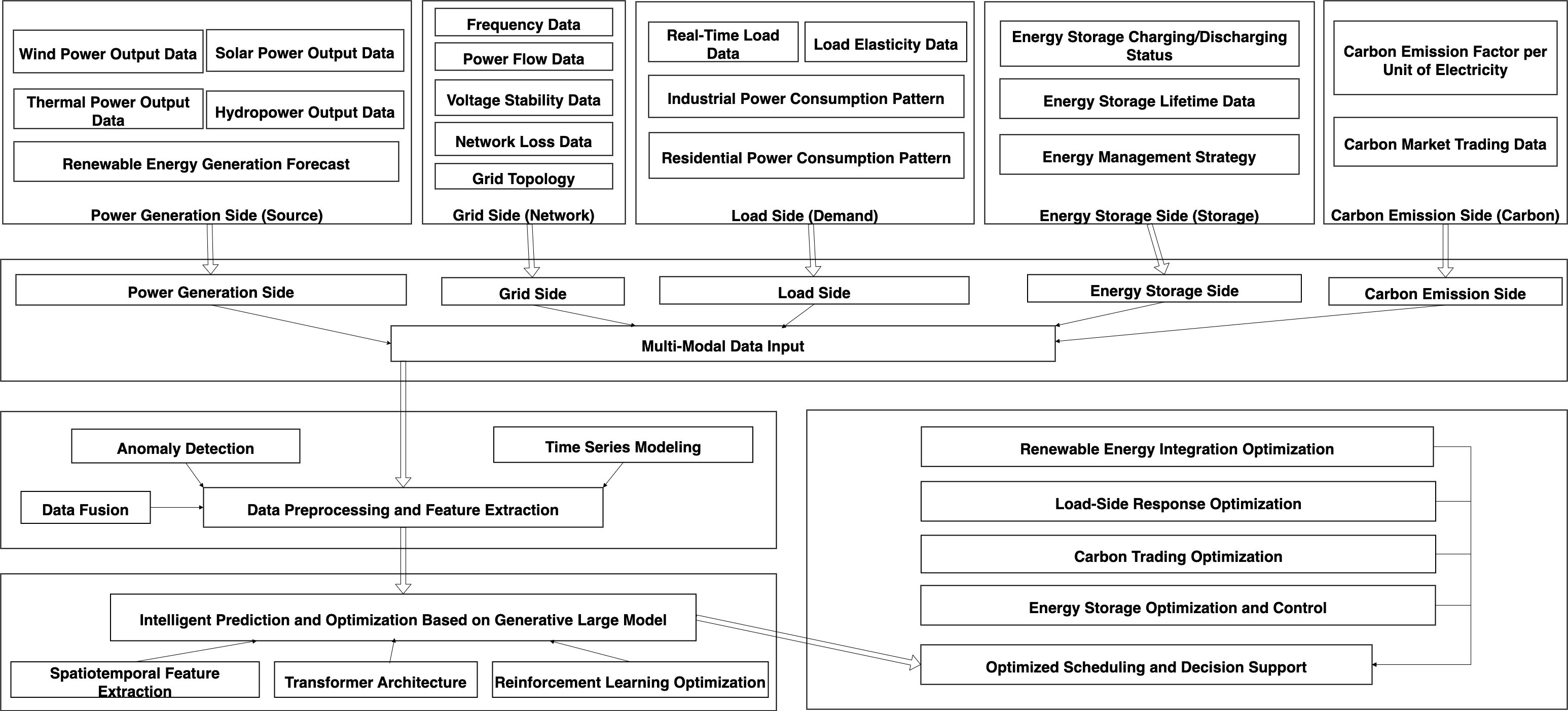}}
\caption{Theoretical Framework for Multi-Element Collaborative Optimization of Source-Grid-Load-Storage-Carbon (SGLSC).}\label{fig2}
\end{figure*}
As mentioned before, the construction of a modern power system requires a holistic approach showing as Fig. \ref{fig2} that integrates five core components: power sources, grid, load, energy storage, and carbon emissions. This integration is essential for achieving the goals of high penetration of renewable energy, stable and secure grid operation, efficient energy utilization, and minimizing carbon emissions. However, traditional power systems often operate these elements in isolation, lacking the necessary collaborative optimization mechanisms to address the challenges posed by fluctuations in renewable energy output, changes in electricity demand, and the management of carbon emissions \cite{10838247}.

GLMs with their powerful data-driven capabilities and pattern recognition features, offer a promising solution for the collaborative optimization of the SGLSC elements. By integrating multi-modal data, GLMs can establish a unified interaction analysis framework, accurately predicting the dynamic interactions between these components, optimizing energy allocation, and enhancing the overall operational efficiency of the system. Specifically:
\begin{itemize}[leftmargin=9px]
    \item In high-renewable grids, GLMs predict wind and solar generation using weather data, optimizing storage operations through reinforcement learning to reduce curtailment and enhance grid stability.
    \item For carbon and power market coordination, GLMs forecast carbon price trends, optimize generation mixes for lower emissions, and refine carbon trading strategies, improving market efficiency and reducing costs for power producers.
    \item During extreme weather events, GLMs analyze meteorological data to predict impacts on renewable output and demand, ensuring resilient grid operations.
    \item In smart urban districts, they optimize photovoltaic, storage, and demand-side responses, balancing economic and environmental benefits for efficient energy management.
\end{itemize}

\section{Conclusion}
GLMs offer a transformative approach to addressing the challenges of renewable energy integration in modern power systems. This paper examines GLMs across key sectors—generation, grid, load, storage, and carbon management—highlighting their potential to improve system stability, efficiency, and sustainability. In generation, GLMs enhance renewable energy forecasting and optimize dispatch strategies, reducing curtailment and improving grid integration. On the grid side, they strengthen situational awareness, fault detection, and risk forecasting through advanced modeling techniques. For load management, GLMs enable dynamic forecasting, demand response, and adaptive strategies, improving resilience and energy optimization. In storage, they optimize dispatch and charging strategies, ensuring system efficiency and stability during extreme conditions. GLMs also support carbon emission forecasting and market optimization, aiding the transition to low-carbon power systems. By providing a unified framework for optimizing SGLSC, GLMs improve overall system efficiency, minimize emissions, and ensure stability. Future research should focus on enhancing GLMs' scalability, interpretability, and integration with emerging technologies to advance smart grid and sustainable energy solutions.

\bibliographystyle{IEEEtran}
\bibliography{IEEEabrv,RefDatabase}

% Generated by IEEEtran.bst, version: 1.14 (2015/08/26)
\begin{thebibliography}{10}
\providecommand{\url}[1]{#1}
\csname url@samestyle\endcsname
\providecommand{\newblock}{\relax}
\providecommand{\bibinfo}[2]{#2}
\providecommand{\BIBentrySTDinterwordspacing}{\spaceskip=0pt\relax}
\providecommand{\BIBentryALTinterwordstretchfactor}{4}
\providecommand{\BIBentryALTinterwordspacing}{\spaceskip=\fontdimen2\font plus
\BIBentryALTinterwordstretchfactor\fontdimen3\font minus \fontdimen4\font\relax}
\providecommand{\BIBforeignlanguage}[2]{{%
\expandafter\ifx\csname l@#1\endcsname\relax
\typeout{** WARNING: IEEEtran.bst: No hyphenation pattern has been}%
\typeout{** loaded for the language `#1'. Using the pattern for}%
\typeout{** the default language instead.}%
\else
\language=\csname l@#1\endcsname
\fi
#2}}
\providecommand{\BIBdecl}{\relax}
\BIBdecl

\bibitem{10252793}
Y.~Tang, Y.~Lu, X.~Yang, G.~Liu, T.~Liu, and T.~Yang, ``“one graph of electricity carbon” spatiotemporal data analysis and management system,'' in \emph{IEEE Power \& Energy Society General Meeting (PESGM)}, 2023, pp. 1--5.

\bibitem{10838247}
P.~Li, Z.~Dai, Y.~Tang, G.~Liu, J.~Hou, Q.~Feng, and Q.~Lin, ``Spatiotemporal data graph modeling and exploration of application scenarios in “power grid one graph”,'' \emph{CSEE Journal of Power and Energy Systems}, pp. 1--12, 2025.

\bibitem{9299496}
B.~Xiao, J.~Wang, Z.~Xiao, G.~Yan, L.~Dong, M.~Wang, and H.~Yang, ``Power source flexibility margin quantification method for multi-energy power systems based on blind number theory,'' \emph{CSEE Journal of Power and Energy Systems}, vol.~9, no.~6, pp. 2321--2331, 2023.

\bibitem{8999526}
S.~Peyghami, P.~Palensky, and F.~Blaabjerg, ``An overview on the reliability of modern power electronic based power systems,'' \emph{IEEE Open Journal of Power Electronics}, vol.~1, pp. 34--50, 2020.

\bibitem{9726814}
S.~De, M.~Bermudez-Edo, H.~Xu, and Z.~Cai, ``Deep generative models in the industrial internet of things: A survey,'' \emph{IEEE Transactions on Industrial Informatics}, vol.~18, no.~9, pp. 5728--5737, 2022.

\bibitem{9721662}
Y.~Cheng, N.~Yu, B.~Foggo, and K.~Yamashita, ``Online power system event detection via bidirectional generative adversarial networks,'' \emph{IEEE Transactions on Power Systems}, vol.~37, no.~6, pp. 4807--4818, 2022.

\bibitem{10166280}
Y.~Chen, Y.~Tang, S.~Zhang, G.~Liu, and T.~Liu, ``Weather sensitive residential load forecasting using neural networks,'' in \emph{IEEE 6th International Electrical and Energy Conference (CIEEC)}, 2023, pp. 3392--3397.

\bibitem{9462413}
M.~A. Hossain, R.~K. Chakrabortty, S.~Elsawah, E.~M. Gray, and M.~J. Ryan, ``Predicting wind power generation using hybrid deep learning with optimization,'' \emph{IEEE Transactions on Applied Superconductivity}, vol.~31, no.~8, pp. 1--5, 2021.

\bibitem{9555586}
M.~Massaoudi, I.~Chihi, H.~Abu-Rub, S.~S. Refaat, and F.~S. Oueslati, ``Convergence of photovoltaic power forecasting and deep learning: State-of-art review,'' \emph{IEEE Access}, vol.~9, pp. 136\,593--136\,615, 2021.

\bibitem{9869310}
W.~Cui, C.~Wan, and Y.~Song, ``Ensemble deep learning-based non-crossing quantile regression for nonparametric probabilistic forecasting of wind power generation,'' \emph{IEEE Transactions on Power Systems}, vol.~38, no.~4, pp. 3163--3178, 2023.

\bibitem{10512990}
G.~Hongxia, L.~Yuan, C.~Lingxuan, W.~Ziqiang, M.~Qian, and L.~Yiming, ``An improved generative adversarial network for extreme scenarios generation,'' in \emph{IEEE 7th Conference on Energy Internet and Energy System Integration (EI2)}, 2023, pp. 1472--1477.

\bibitem{9583953}
J.~Guan, H.~Tang, J.~Wang, J.~Yao, K.~Wang, and W.~Mao, ``A gan-based fully model-free learning method for short-term scheduling of large power system,'' \emph{IEEE Transactions on Power Systems}, vol.~37, no.~4, pp. 2655--2665, 2022.

\bibitem{10632043}
Z.~Dai, S.~Liang, Y.~Tang, J.~Tan, G.~Liu, Q.~Feng, and X.~Li, ``Efficient state estimation through rapid topological analysis based on spatiotemporal graph methodology,'' \emph{IEEE Open Access Journal of Power and Energy}, vol.~11, pp. 396--409, 2024.

\bibitem{9337223}
G.~Cheng, Y.~Lin, Y.~Chen, and T.~Bi, ``Adaptive state estimation for power systems measured by pmus with unknown and time-varying error statistics,'' \emph{IEEE Transactions on Power Systems}, vol.~36, no.~5, pp. 4482--4491, 2021.

\bibitem{9321176}
M.~R.~R. Mojumdar, J.~M. Cano, and G.~A. Orcajo, ``Estimation of impedance ratio parameters for consistent modeling of tap-changing transformers,'' \emph{IEEE Transactions on Power Systems}, vol.~36, no.~4, pp. 3282--3292, 2021.

\bibitem{8667290}
Z.~Pan, W.~Yu, X.~Yi, A.~Khan, F.~Yuan, and Y.~Zheng, ``Recent progress on generative adversarial networks (gans): A survey,'' \emph{IEEE Access}, vol.~7, pp. 36\,322--36\,333, 2019.

\bibitem{9119397}
P.~Tang, K.~Peng, J.~Dong, K.~Zhang, and S.~Zhao, ``Monitoring of nonlinear processes with multiple operating modes through a novel gaussian mixture variational autoencoder model,'' \emph{IEEE Access}, vol.~8, pp. 114\,487--114\,500, 2020.

\bibitem{10176302}
M.~Gao, J.~Yu, Z.~Yang, and J.~Zhao, ``Physics embedded graph convolution neural network for power flow calculation considering uncertain injections and topology,'' \emph{IEEE Transactions on Neural Networks and Learning Systems}, vol.~35, no.~11, pp. 15\,467--15\,478, 2024.

\bibitem{9555209}
S.~Bond-Taylor, A.~Leach, Y.~Long, and C.~G. Willcocks, ``Deep generative modelling: A comparative review of vaes, gans, normalizing flows, energy-based and autoregressive models,'' \emph{IEEE Transactions on Pattern Analysis and Machine Intelligence}, vol.~44, no.~11, pp. 7327--7347, 2022.

\bibitem{8881675}
Y.~Tang, C.-W. Ten, and K.~P. Schneider, ``Inference of tampered smart meters with validations from feeder-level power injections,'' in \emph{IEEE Innovative Smart Grid Technologies - Asia (ISGT Asia)}, 2019, pp. 2783--2788.

\bibitem{9729853}
H.~Yang, K.~Zhang, and A.~Tang, ``Risk assessment of main electrical connection in substation with regional grid safety constraints,'' \emph{IEEE Access}, vol.~10, pp. 27\,750--27\,758, 2022.

\bibitem{10379576}
Z.~Zhang, S.~Liao, Y.~Sun, J.~Xu, D.~Ke, B.~Wang, R.~Chen, and Y.~Jiang, ``A parallel-type load damping factor controller for frequency regulation in power systems with high penetration of renewable energy sources,'' \emph{Journal of Modern Power Systems and Clean Energy}, vol.~12, no.~4, pp. 1019--1030, 2024.

\bibitem{9319813}
J.~Wang, X.~Chen, F.~Zhang, F.~Chen, and Y.~Xin, ``Building load forecasting using deep neural network with efficient feature fusion,'' \emph{Journal of Modern Power Systems and Clean Energy}, vol.~9, no.~1, pp. 160--169, 2021.

\bibitem{9698190}
P.~Manohar and C.~R. Atla, ``Development of predictive reliability model of solar photovoltaic system using stochastic diffusion process for distribution system,'' \emph{IEEE Journal on Emerging and Selected Topics in Circuits and Systems}, vol.~12, no.~1, pp. 279--289, 2022.

\bibitem{7741548}
J.~Yan, H.~Zheng, and N.~Lu, ``Temperature-load sensitivity study for adjusting miso day-ahead load forecast,'' in \emph{IEEE Power and Energy Society General Meeting (PESGM)}, 2016, pp. 1--5.

\bibitem{10368142}
W.~Huang, N.~Zhang, C.~Kang, M.~Li, and M.~Huo, ``From demand response to integrated demand response: review and prospect of research and application,'' \emph{Protection and Control of Modern Power Systems}, vol.~4, no.~2, pp. 1--13, 2019.

\bibitem{10166065}
J.~Luan, Y.~Tang, Y.~Lit, G.~Li, and S.~Zhang, ``A data trust model based on alarms considering evs participating in market transactions,'' in \emph{6th International Conference on Energy, Electrical and Power Engineering (CEEPE)}, 2023, pp. 1488--1493.

\bibitem{10423192}
S.~Chakraborty, S.~Bera, S.~Kar, and S.~R. Samantaray, ``Ensuring long term sustainability in networked microgrids through intelligent load management and priority-based power transfer scheme,'' \emph{IEEE Transactions on Power Delivery}, vol.~39, no.~3, pp. 1386--1398, 2024.

\bibitem{8978655}
N.~Huang, W.~Wang, S.~Wang, J.~Wang, G.~Cai, and L.~Zhang, ``Incorporating load fluctuation in feature importance profile clustering for day-ahead aggregated residential load forecasting,'' \emph{IEEE Access}, vol.~8, pp. 25\,198--25\,209, 2020.

\bibitem{10875413}
Q.~Sun, Z.~Zhao, M.~Fan, and Q.~Ma, ``Electricity market price forecasting model based on an improved long short-term memory network,'' in \emph{4th International Conference on Energy, Power and Electrical Engineering (EPEE)}, 2024, pp. 903--906.

\bibitem{8683792}
H.~Rashid, V.~Stankovic, L.~Stankovic, and P.~Singh, ``Evaluation of non-intrusive load monitoring algorithms for appliance-level anomaly detection,'' in \emph{IEEE International Conference on Acoustics, Speech and Signal Processing (ICASSP)}, 2019, pp. 8325--8329.

\bibitem{6246270}
C.~Chen, Y.~Liu, and H.-Z. Huang, ``Optimal load distribution for multi-state systems under selective maintenance strategy,'' in \emph{International Conference on Quality, Reliability, Risk, Maintenance, and Safety Engineering}, 2012, pp. 436--442.

\bibitem{7491338}
Y.~Zhang, N.~Rahbari-Asr, J.~Duan, and M.-Y. Chow, ``Day-ahead smart grid cooperative distributed energy scheduling with renewable and storage integration,'' \emph{IEEE Transactions on Sustainable Energy}, vol.~7, no.~4, pp. 1739--1748, 2016.

\bibitem{10398613}
S.~Zhai, Y.~Tang, X.~Zhu, H.~Ye, X.~Lu, and G.~Liu, ``Graph computing-based energy storage profit optimization considering market participant,'' in \emph{6th International Conference on Power and Energy Applications (ICPEA)}, 2023, pp. 307--312.

\bibitem{8993691}
Z.~Shi, W.~Wang, Y.~Huang, P.~Li, and L.~Dong, ``Simultaneous optimization of renewable energy and energy storage capacity with the hierarchical control,'' \emph{CSEE Journal of Power and Energy Systems}, vol.~8, no.~1, pp. 95--104, 2022.

\bibitem{10379573}
J.~Li, Z.~Fang, Q.~Wang, M.~Zhang, Y.~Li, and W.~Zhang, ``Optimal operation with dynamic partitioning strategy for centralized shared energy storage station with integration of large-scale renewable energy,'' \emph{Journal of Modern Power Systems and Clean Energy}, vol.~12, no.~2, pp. 359--370, 2024.

\bibitem{10512421}
S.~Zhao, Y.~Tang, Z.~Huang, Q.~Li, G.~Liu, and S.~Zhang, ``Research on high-frequency spatiotemporal electrical carbon intensity,'' in \emph{IEEE 7th Conference on Energy Internet and Energy System Integration (EI2)}, 2023, pp. 3055--3060.

\bibitem{9721566}
G.~Liu, J.~Liu, J.~Zhao, J.~Qiu, Y.~Mao, Z.~Wu, and F.~Wen, ``Real-time corporate carbon footprint estimation methodology based on appliance identification,'' \emph{IEEE Transactions on Industrial Informatics}, vol.~19, no.~2, pp. 1401--1412, 2023.

\bibitem{9056463}
H.~He, Z.~Luo, Q.~Wang, M.~Chen, H.~He, L.~Gao, and H.~Zhang, ``Joint operation mechanism of distributed photovoltaic power generation market and carbon market based on cross-chain trading technology,'' \emph{IEEE Access}, vol.~8, pp. 66\,116--66\,130, 2020.

\end{thebibliography}
\end{document}